\documentclass[%
 reprint,
 amsmath,amssymb,
prb,
,superscriptaddress]{revtex4-2}   

\usepackage{graphicx}
\usepackage{dcolumn}
\usepackage{bm}
\usepackage{color}
\usepackage[dvipsnames, svgnames, x11names]{xcolor} 
\usepackage{amsmath}
\usepackage[colorlinks=true,
           linkcolor={Blue},
           anchorcolor={Blue},
           citecolor={Blue},
           urlcolor={Blue}
           ]{hyperref} 

\usepackage{changes} 
\usepackage{lipsum}
\definechangesauthor[name={An An}, color=orange]{An}
\definechangesauthor[name={Hong Hong}, color=black]{Hong}

\usepackage{booktabs}
\usepackage{diagbox} 
\usepackage{multirow} 
\usepackage{makecell} 
\usepackage{longtable} 

\usepackage{CJK}
\usepackage{indentfirst}
\usepackage{cases}

\begin{document}

\preprint{APS/123-QED}

\title{Bistability in dissipatively coupled cavity magnonics}

\author{H. Pan}
\email{hpan19@fudan.edu.cn}
\affiliation{State Key Laboratory of Surface Physics, Department of Physics, Fudan University, Shanghai 200433, People's Republic of China}
\affiliation{Department of Physics and Astronomy, University of Manitoba, Winnipeg, Canada R3T 2N2}

\author{Y. Yang}
\affiliation{Department of Physics and Astronomy, University of Manitoba, Winnipeg, Canada R3T 2N2}

\author{Z.H. An}
\email{anzhenghua@fudan.edu.cn}
\affiliation{State Key Laboratory of Surface Physics, Department of Physics, Fudan University, Shanghai 200433, People's Republic of China}
\affiliation{Institute of Nanoelectronic Devices and Quantum Computing, Fudan University, Shanghai 200433, People's Republic of China}
\affiliation{Shanghai Qi Zhi Institute, 41th Floor, AI Tower, No. 701 Yunjin Road, Xuhui District, Shanghai, 200232, People's Republic of China}

\author{C.-M. Hu}
\email{hu@physics.umanitoba.ca}
\affiliation{Department of Physics and Astronomy, University of Manitoba, Winnipeg, Canada R3T 2N2}

\date{\today}

\begin{abstract}
  Dissipative coupling of resonators arising from their cooperative dampings to a common reservoir induces intriguingly new physics such as energy level attraction.
  In this study, we report the nonlinear properties in a dissipatively coupled cavity
  magnonic system. A magnetic material YIG~(yttrium iron garnet) is placed at the magnetic field node of a Fabry-Perot-like microwave cavity such that the magnons and cavity photons are dissipatively coupled. Under high power excitation, a nonlinear effect is observed in the transmission spectra, showing bistable behaviors. The observed bistabilities are manifested as clockwise, counterclockwise, and butterfly-like hysteresis loops with different frequency detuning. The experimental results are well explained as a Duffing oscillator dissipatively coupled with a harmonic one and the required trigger condition for bistability could be determined quantitatively by the coupled oscillator model. Our results demonstrate that the magnon damping has been suppressed by the dissipative interaction, which thereby reduces the threshold for conventional magnon Kerr bistability. This work sheds light upon potential applications in developing low power nonlinearity devices, enhanced anharmonicity sensors and for exploring the non-Hermitian physics of cavity magnonics in the nonlinear regime.
\end{abstract}

\maketitle


\section{\label{sec:level1}Introduction}

Nonlinearities are ubiquitous phenomena in various physics fields. For instance, Kerr nonlinearity and resonant two-level nonlinearity~\cite{camara2015optical}, leading to anharmonicities, have been widely investigated in the context of optics~\cite{lugiato2015nonlinear}. One signature of nonlinear dynamics is bistability for a given input exceeding a threshold power and manifests itself as a foldover effect~\cite{landau2000mechanics,mckinstry1985low,Gui2009Direct,Gui2009Foldover,Janantha2017Foldover}. The effects of nonlinearity have technological implications in sophisticated optical devices for controlling light with light~\cite{gibbs2012optical}, novel magnetic data storage devices~\cite{thirion2003switching}, switches with bistable metamaterial~\cite{Bilal2017Bistable} as well as developed mechanical devices for emergent applications like energy harvesting~\cite{Cottone2009Nonlinear,harne2013review}. 

  Hybrid quantum systems, which have aroused tremendous interest for application in quantum information processing~\cite{xiang2013hybrid,kurizki2015quantum}, have the potential to push the development of the realm of nonlinearity. Among various hybrid subsystems, cavity magnonics appears to be an exceptional candidate (see, e.g.,~\cite{huebl2013high,tabuchi2014hybridizing,zhang2014strongly,bai2015spin,zhang2015magnon,tabuchi2015coherent,chumak2015magnon,zhang2016cavity,osada2016cavity,Bai2017Cavity,zhang2017observation,Li2018Magnon,Zhang2019Quantum,Nair2020Deterministic,Wang2018Bistability,nair2020nonlinear}), which utilizes ferrimagnetic materials like YIG~(yttrium iron garnet) to create collective spin excitations. This cavity magnonics system has resulted in a variety of semiclassical and quantum phenomena including, cavity-magnon polaritons~\cite{Cao2015Exchange,Yao2015Theory,Hyde2017Linking}, magnon bistability~\cite{Wang2018Bistability,Wang2016Magnon,Hyde2018Direct}, bidirectional microwave-optical conversion mediated by ferromagnetic magnons~\cite{Hisatomi2016Bidirectional}, magnon dark mode~\cite{zhang2015magnon}, synchronization via spin-photon coupling~\cite{Grigoryan2018Synchronized}, and non-Hermitian exotic properties~\cite{Harder2017Topological,zhang2017observation,Wang2018Magnon}. Recent work has attempted to utilize the photon-magnon coupling freedom to tune the nonlinearity. The magnon-polariton bistability has been successfully observed in a coherent hybrid system~\cite{Wang2018Bistability} and the bistable behavior could also be directly measured with the coupled cavity being pumped~\cite{Hyde2018Direct}. However, in these reported works the photon-magnon are typically coherently coupled, and their coherent coupling enhances the damping of magnon states~\cite{yao2019microscopic}. Due to the cubic dependence of the threshold for nonlinearity on the effective damping of magnon~\cite{Hyde2018Direct}, the coherent coupling of cavity-magnon hybrid system therefore increases the nonlinearity threshold which may impede the device application of nonlinearity.

Coherent coupling originates from the direct spatial overlap between photon and magnon modes and forms hybrid states with repulsed energy levels and attracted damping rates. In contrast, a dissipative form of magnon-photon interaction~\cite{Harder2018Level} requires no direct mode overlap and has been demonstrated to cause level attraction and a damping repulsion effect\cite{Bhoi2019Abnormal,Yang2019Control,Rao2019Level}. This dissipative coupling is indirect as it is mediated through a shared reservoir, resulting in an imaginary spin-photon coupling strength. Dissipatively coupled systems have important applications such as nonreciprocal transport~\cite{wang2019nonreciprocity}, enhanced sensing~\cite{Nair2021Enhanced}, and non-Hermitian singularities~\cite{ashida2020non,Yang2020Unconventional}. So far, however, reported work with dissipative coupling have been focused on the linear regime and the nonlinearity with dissipative coupling have not been experimentally examined, except a theoretical prediction of a lower threshold if the (imaginary) coupling strength is sufficiently large in an anti-PT regime~\cite{Nair2021Ultralow}.

Inspired by this, we study a dissipatively coupled cavity-magnon system where a YIG sphere is imbedded in a Fabry-Perot-like cavity. The coupled system is directly pumped through the cavity and the pumping power is sufficiently high to create the nonlinear effect. We experimentally demonstrate that, in the dissipatively coupled nonlinear system, the threshold power for bistability is lower than the corresponding bound in coherent scenario. This is due to the suppression of magnon damping on-resonance induced by dissipative coupling. In addition, such a dissipatively coupled hybridized system results in different magnetic field and power dependent bistable behaviors as the magnon frequency is detuned for on and off cavity resonant frequency. By introducing the model based on a Duffing oscillator dissipatively coupled with a harmonic oscillator, the experimental results can be well explained. This method allows us to determine the required experimental condition for triggering bistable behavior in a dissipatively coupled system. Our work shows that dissipative coupling in a hybrid system with a nonlinear effect may be utilized to engineer a lower power threshold for nonlinearity and enhanced anharmonicity sensors. 

\section{THEORETICAL model}

We begin with a model considering the nonlinear Kerr effect of the magnon excitation in a YIG sphere which dissipatively interacts with the cavity photons. Here the cavity is directly pumped. This dissipatively coupled nonlinear system is characterized by a Hamiltonian (the details shown in Appendix \ref{Hamiltonian of the coupled hybrid system} and with reduced units $\hbar\equiv 1$):
\begin{center}
\begin{eqnarray}
  &&\hspace{-1.5em}H_{tot} = {\omega _c}{a^ + }a + {\omega _m}{b^ + }b + K{b^ + }b{b^ + }b + i\Gamma \left( {{a^ + }b + a{b^ + }} \right)  \nonumber\\
  &&\hspace{-1.5em}~~~~~~~~~~~~~~~~~~~~+{\Omega _d}\left( {{a^ + }{e^{ - i\omega t}} + a{e^{i\omega t}}} \right),
\end{eqnarray}
\end{center}
where $a^+$~$(a)$ and $b^+$~$(b)$ correspond to the creation (annihilation) operators of the cavity photons at frequency $\omega _c$ and of the Kittel-mode magnons at $\omega _m$, respectively. Here, the Kerr effect of magnons term $Kb^+bb^+b$ originates from magnetocrystalline anisotropy in the YIG material~\cite{Wang2018Bistability}, in which $K$ is the Kerr coefficient and is positive in our experiment specifically (see the Appendix \ref{Hamiltonian of the coupled hybrid system}). $\varGamma$ is the dissipative coupling strength between the cavity photon and the magnon via their common reservoir. The last term in the above equation describes that the cavity is pumped by the oscillating microwave field, with an amplitude $\Omega _d$~\cite{agarwal2010electromagnetically}.

By adopting the Heisenberg-Langevin approach~\cite{walls2008quantum}, the dynamics of the coupled cavity photon-magnon hybrid nonlinear system can be described by the equations:
\begin{eqnarray}
  &&\hspace{-2em}\textcolor{black}{\frac{da}{dt} =  - i({\omega _c} - i\kappa )a + \Gamma b - i{\Omega _d}},\\
  &&\hspace{-2em}\textcolor{black}{\frac{db}{dt} =  - i\left({{\omega _m} - i\gamma} \right)b-i(Kbb^+b}   
   \textcolor{black}{+Kb^+bb) + \Gamma a.}
\end{eqnarray}
Under the mean-field approximation~\cite{Nair2021Enhanced}, the higher-order expectations can be decoupled, so that the nonlinear term $\left \langle b^+bb\right \rangle$ and $\left \langle bb^+b\right \rangle$ can be simplified as $|b|^2b$, then the dynamics of the dissipatively coupled magnon-photon system follows as,
\begin{eqnarray}
  &&\hspace{-2em}\frac{{da}}{{dt}} =  - i({\omega _c} - i\kappa )a + \Gamma b - i{\Omega _d},\label{dynamic equation of a}\\
  &&\hspace{-2em}\frac{{db}}{{dt}} =  - i\left( {\omega _m} + 2K{|b|^2} - i\gamma \right)b + \Gamma a,
\label{dynamic equation of b}
\end{eqnarray}
where $\kappa =\kappa _i+\kappa _e$ is the total damping of the cavity mode, with $\kappa _i$~$(\kappa _e)$ being the intrinsic (external) damping of the cavity mode. $\gamma $ is the damping rate of the Kittel-mode. Suppose the photon and magnon mode \textcolor{black}{have time dependence of $e^{-i\omega t}$, then Eqs.~(\ref{dynamic equation of a}) and (\ref{dynamic equation of b}) can be simplified as }
\begin{eqnarray}
  \left( {\omega  - {\omega _c} + i\kappa } \right)a - i\Gamma b = {\Omega _d},\label{eqnsolve1}\\
  \left( {\omega  - {\omega _m} - 2K{{\left| b \right|}^2} + i\gamma} \right)b = i\Gamma a.\label{eqnsolve2}
\end{eqnarray}
Eq.~(\ref{eqnsolve1}) can be expressed as $a=(i\varGamma b+\varOmega _d)/(\omega -\omega _c+i\kappa)$. By substituting such an expression into Eq.~(\ref{eqnsolve2}), we get
\begin{equation}
  |b| \left[ \gamma {_m}'-i\left(\delta \omega{_m}'-2K|b|^{2}\right) \right]=\frac{\varGamma \varOmega _d}{\delta \omega _c+i\kappa}. \label{nonlinear_b_field}
\end{equation}
Here, $\delta \omega {_m}'=\delta \omega _m+\eta \delta \omega _c$ with $\delta \omega_m=\omega -\omega _m$, $\delta \omega_c=\omega -\omega_c$ denotes effective frequency shift, and 
\begin{equation}
 \gamma {_m}'=\gamma -\eta \kappa,
 \label{damping of magnon}  
\end{equation}
denotes the effective damping of the magnon for the dissipative coupling system, where the negative sign shows the suppression effect of the magnon damping. We note that for the coherent coupling, the sign is positive representing the enhancement effect of the magnon damping~\cite{Wang2018Bistability,Hyde2018Direct}. The coefficient $\eta=\Gamma ^{2}/(\delta \omega _c^{2}+\kappa ^{2})$ stands for the transfer efficiency of the excitation power $P$ from the input port into the magnon system, and $\eta$ is dependent on the dissipative coupling strength $\Gamma$, the frequency detuning $\delta \omega _c$, and the total damping $\kappa $ of the cavity.

Taking the squared modulus of Eq.~(\ref{nonlinear_b_field}) and defining $\varDelta _m\equiv 2K|b|^{2}$, which is the shift of the magnon frequency~\cite{Wang2016Magnon}, we have
\begin{equation}
  \varDelta _m \left[ {\gamma _m'}^2+{\left(\delta \omega_m'-\varDelta _m\right)}^{2} \right]=2\eta K\Omega_d^2. \label{nonlinear_b_power}
\end{equation}
  Equation~(\ref{nonlinear_b_power}) has a similar form to the uncoupled Duffing oscillator~\cite{landau2000mechanics}, except that $\gamma_m'$ and $\delta \omega_m'$ are the effective damping and frequency shift of the magnon, respectively, which result from dissipative coupling with the cavity photon. The right term of Eq.~(\ref{nonlinear_b_power}) denotes the effective drive field of the magnon via dissipative interaction with the cavity photon as the field directly pumps the cavity rather than the YIG sphere.

\subsection{Bistability and the transition point}
Equation~(\ref{nonlinear_b_power}) describes an oscillator with cubic nonlinearity, called the Duffing equation. As for this equation, there are three real roots for a finite range of frequencies when $\Omega_d$ exceeds a specific value called the critical field $\Omega_t$. Among the three roots, one is unstable and the other two are stable, called bistability, which is a signature of the anharmonic oscillator. We will focus on the two boundaries (hereafter termed as up border and down border) inbetween which bistability occurs. As there are abrupt transitions between two stable states at up and down borders, the transition points of bistability are determined by the condition $d\Omega _d/d\varDelta _m=0$, i.e.,
\begin{equation}
  3\varDelta _m^{2}-4\delta \omega _m'\varDelta_m+{\delta \omega _m'}^{2}+{\gamma _m'}^{2} =0.
  \label{quadratic equation for boarder}
\end{equation}
According to the root discriminant of the quadratic equation, when $4\delta {\omega _m'}^{2}-12{\gamma _m'}^{2}=0$ , i.e.,
\begin{equation}
  \delta \omega _m'=-\sqrt3\gamma_m',
  \label{critical solution for qudratic equation}
\end{equation}
there is only one root of Eq.~(\ref{quadratic equation for boarder}), which corresponds to the critical condition of bistable behavior.\\

When $4\delta {\omega _m'}^{2}-12{\gamma _m'}^{2}>0$, there are two real roots which correspond to the up border and down border of the bistability. By adopting the approximation $\sqrt{3}\gamma_m'\ll \delta\omega _m'$ to solve Eq.~(\ref{quadratic equation for boarder}), then the upper border satisfies
\begin{equation}
  \delta \omega_m=3\sqrt[3]{\frac{\eta KS^2P}{2}}-\eta \delta \omega_c,~\text{(up)}
  \label{upper border}
\end{equation}
and the down border satisfies 
\begin{equation}
  \delta \omega_m=\frac{2\eta KS^2P}{{(\gamma-\eta \kappa)}^{2}}-\eta \delta \omega _c,~\text{(down)}
\label{down boarder}
\end{equation}
where $\Omega _d=S\sqrt{P}$, with $S$ describing the conversion efficiency from input power $P$ to the field $\Omega_d$ driving the cavity resonance mode. The magnitude of $S$ depends on the frequency, phenomenologically introduced external loss of the cavity field, and the loss in the cable which connects the device.
Based on Eqs.~(\ref{upper border}) and (\ref{down boarder}), we can come to conclusions that both the upper border and down border have input power dependence, with the upper border satisfying $\delta \omega_m\propto {P}^{1/3}$ compared with down border satisfying $\delta \omega_m\propto P$. These dependencies are different from those of coherently coupled anharmonic oscillators~\cite{Hyde2018Direct}, where the upper border has $\delta \omega_m\propto P$ dependence and the down border $\delta \omega_m\propto {P}^{1/3}$ dependence. The difference arises from that the magnon shifts to higher frequencies with negative Kerr coefficient $K$ while that of our work shifts to lower frequencies with positive Kerr coefficient. In fact, the two cases can be unified where the border with a large frequency shift has a $P$ dependence and the border with a small frequency shift has a $P^{1/3}$ dependence.

The above discussion is based on tuning the magnetic field to obtain bistability, called field bistability, and its transition point has power dependence. On the other hand, we can adjust input power to obtain bistability, called power bistability, and its transition point has magnetic field dependence. According to Eqs.~(\ref{upper border}) and (\ref{down boarder}), we can get borders of power bistability by regarding magnetic field $H$ as an independent variable
\begin{eqnarray}
  &&\hspace{-2em}P=\frac{2(\delta \omega_m+\eta\delta \omega_c)^3}{27\eta KS^2}, ~\text{(up)} 
  \label{up power stability}\\
  &&\hspace{-2em}P=\frac{(\delta\omega_m+\eta\delta\omega_c)(\gamma-\eta\kappa^2)}{2\eta KS^2}.~\text{(down)}
  \label{down} 
\end{eqnarray}
As the magnon resonance detuning $\delta \omega_m=\gamma _0 \delta H$ with definition $\delta H=H-H_r$ ($H_r$ is the resonant magnetic field and $\gamma _0$ is gyromagnetic ratio), the up and down border of power bistability have respective $|{\delta H}| ^{3}$ and $|\delta H|$ dependence when up-sweeping and down-sweeping power as shown in Eqs.~(\ref{up power stability}) and (\ref{down}), respectively. 

\subsection{Critical condition of bistability}
When Eq.~(\ref{quadratic equation for boarder}) has only one real solution, the bistability vanishes because the two transition points collapse to one point. The critical driving field of the cavity corresponds to the threshold beyond which the bistability appears. Thus, by substituting the critical condition of bistability shown in Eq.~(\ref{critical solution for qudratic equation}) and the only solution $\varDelta _m=2/3\delta \omega _m$ into Eq.~(\ref{nonlinear_b_power}), the critical field $\Omega_t$ (critical power $P_t$) of field bistability is obtained: 

\begin{subequations}
  \begin{align}
  \Omega_t^2=\frac{4\sqrt{3}}{9K}\frac{\gamma _m'^3}{\eta},\label{omega_t}\\
 P_t=\frac{4\sqrt{3}}{9KS^2}\frac{\gamma _m'^3}{\eta} \label{P_t}.
  \end{align}
\end{subequations}
These equations imply that the threshold has cubic dependence on the effective damping of magnon $\gamma_m'$, which is similar to that of coherently coupled nonlinear system~\cite{Hyde2018Direct}. The threshold of nonlinearity for the hybrid system have similar dependence with that of uncoupled ferrimagnetic resonance which has cubic dependence upon the damping of magnon~(see Table \ref{Threshold}).
Next, we compare the threshold value for the dissipatively coupled system with that of the coherently coupled system. For this purpose, the threshold for a generic two-mode system involving both coherent and dissipative coupling is generated by substituting $\gamma _m'=\gamma+\eta\kappa$ with $\eta={(g+i\Gamma)^2}/{(\delta\omega_c^2+\kappa^2)}$ into Eqs.~(\ref{omega_t}) and (\ref{P_t}) where $g$ is the coherent coupling strength. Supposing scenarios $\Gamma=0$ and $g = 0$, we will get the threshold for coherently coupled and dissipatively coupled systems, respectively. By comparing the thresholds in relation to the nature of coupling~(setting $\delta\omega_c=0$), we have

\begin{equation}
  \frac{{({\Omega_t}^{d})}^2}{{({\Omega_t}^{c})}^2}=\frac{{P_t}^{d}}{{P_t}^{c}}=\frac{g^{2}}{\Gamma ^{2}}\left|\frac{1-\Gamma^{2}/\kappa \gamma}{1+g^2/\kappa\gamma}\right|^{3},
  \label{low threshold} 
\end{equation}
where we have assumed that the two kinds of coupling systems have the same damping coefficient and ${\Omega_t}^{d}~(P_t^d)$, ${\Omega_t}^{c}~(P_t^c)$ denote the critical field (power) of bistability for dissipative and coherent coupling respectively. 

\begin{table}[!htbp]
	\centering
	\caption{Thresholds of the ferrimagnetic materials in the coupled and uncoupled scenarios.}

    \begin{ruledtabular}
	\begin{tabular}
    {lccp{2cm}p{2cm}p{2.8cm}}
    
    \textbf{sample}              &\textbf{threshold}    &\textbf{effective damping}\footnote{Effective damping of magnon of each system.}    \\
		\hline  
    \vspace{1.1mm}
		YIG sphere~\cite{gottlieb1960nonlinear}    & $\gamma_0 ^2h_t^2=\frac{8\sqrt{3}}{9\chi }{\gamma}^3\propto \gamma^3$\footnote{Here, $h_t$ is the critical drive magnetic field, and $\chi$ is related to crystalline anisotropy~\cite{gottlieb1960nonlinear}.}   &$\gamma=\gamma _0 \Delta H$ \footnote{$\Delta H$ is the linewidth of the ferromagnetic resonance.} \\   
    \vspace{1.1mm}              
		Py film~\cite{Gui2009Foldover}    &$\gamma_0 ^2h_{t}^2=\frac{16\sqrt{3}}{9\gamma _0 M_s}{\gamma}^3\propto \gamma^3$       &$\gamma=\gamma_0\Delta H$ \\
    \vspace{1.1mm}
   YIG~(Coh.)\footnote{YIG sphere is placed in a microwave cavity, and they interact coherently.}~\cite{Hyde2018Direct}  &$\Omega _t^2=\frac{4\sqrt{3}}{9K\eta}\gamma _m'^3\propto \gamma_m'^3$   &$\gamma _m'=\gamma+\eta \kappa$  \\
   YIG~(Dis.)\footnote{YIG sphere is placed in a microwave cavity, and they interact dissipatively.} &$\Omega _t^2=\frac{4\sqrt{3}}{9K\eta}\gamma _m'^3\propto \gamma_m'^3$   &$\gamma _m'=\gamma-\eta \kappa$        \\
	\end{tabular}
\end{ruledtabular}
  \label{Threshold}
\end{table}

Notably, the expression of Eq.~(\ref{low threshold}) is always less than $1$ for nonzero $\Gamma$ and $g$, revealing a consistently lower threshold in the dissipatively coupled system than that in the coherently coupled system. In order to make a fair comparison, we assume identical magnitudes of coupling strength, i.e., $\Gamma/2\pi=g/2\pi=21$~MHz. Then, by substituting the value of $\gamma/2\pi=5.1$~MHz and $\kappa/2\pi=126$~MHz into Eq.~(\ref{low threshold}), the dissipative-coherent threshold ratio becomes a finite value of $0.0064$, implying that the dissipative coupling may lead to very low power threshold of the bistability. Such a low threshold arises intrinsically from the suppression of the magnon damping on-resonance [see Eq.~(\ref{damping of magnon})] and the cubic dependence of the threshold upon the effective magnon damping~[indicated by Eqs.~(\ref{omega_t}) and (\ref{P_t})].

On the other hand, we can obtain the requirement to observe the power bistability. The only one root of Eq.~(\ref{quadratic equation for boarder}) described by Eq.~(\ref{critical solution for qudratic equation}) corresponds to the critical magnetic field to generate power bistability. Combining the relation $\delta\omega_m=\gamma _0\delta H$ and Eq.~(\ref{critical solution for qudratic equation}), the critical magnetic field is given by
\begin{equation}
  \delta H=\frac{1}{\gamma _0}[-\eta\delta\omega_c-\sqrt{3}(\gamma-\eta \kappa)].
  \label{critical condition for field}
\end{equation}
Compared with the critical magnetic field condition $H-H_r=\sqrt{3}\gamma/ \gamma_0$ of uncoupled magnetic systems~\cite{heinrich1985fmr,Gui2009Foldover}, the extra term of the magnetic field shift $-\sqrt{3}\eta \kappa/{\gamma _0}$ in Eq.~(\ref{critical condition for field}) results from the effective damping of the magnon resonance near $\omega=\omega_c$~\cite{bai2015spin}. The factor $-\eta\delta \omega_c/ \gamma_0$ represents the additional resonance shift of the magnon which~arises from the interaction between magnon and cavity.

\subsection{Transmission spectra with bistability}
The bistability can be detected experimentally via microwave transmission spectra of the cavity. In this section, we show the magnon frequency shift $\varDelta _m$~(due to the Kerr nonlinearity) is observed in the transmission spectra of the cavity. By considering that the cavity mode couples with the input energy from the port, the dynamic equation Eq.~(\ref{dynamic equation of a}) can be rewritten as
\begin{equation}
  \frac{{da}}{{dt}} =  - i({\omega _c} - i\kappa )a + \Gamma b + \sqrt{\kappa_e}c_{in},\label{a input}
\end{equation}
where $c_{in}$ is the input field. From Eq.~(\ref{eqnsolve2}), the amplitude  of the cavity field can be derived
\begin{equation}
  b=\frac{i\Gamma a}{\omega -\omega _m-\varDelta _m+i\gamma}.
  \label{solution of b}
\end{equation}
According to the input-output theory~\cite{walls2008quantum}, the relation of input-output field can be described as
\begin{equation}
  c_{out}-c_{in}=-\sqrt{\kappa_e}a,
  \label{input output relaiton}
\end{equation}
where $c_{out}$ is the output field.
Combining Eqs.~(\ref{a input}), (\ref{solution of b}), (\ref{input output relaiton}), with the definition of transmission coefficient $S_{21}={c_{out}}/{c_{in}}$, we can obtain the transmission coefficient
\begin{equation}
 S_{21}=1+\frac{\kappa _e}{i(\omega-\omega_c)-\kappa -\Gamma ^{2}/[i(\omega-\omega _m -\varDelta _m)-\gamma] }.
 \label{tansmission equation}
\end{equation}
The transmission of nonlinear dissipatively coupled system in Eq.~(\ref{tansmission equation}) can be reduced to that of two coupled linear oscillators if we perform the transformation $\omega_m+\varDelta _m\rightarrow \omega _m $.~Here, $\varDelta _m$ is the nonlinear magnon resonance frequency shift which is the solution of the Duffing equation as shown in Eq.~(\ref{nonlinear_b_power}). The nonlinear effect can be observed through cavity transmission due to the interaction between the cavity photon and magnon. 
\section{EXPERIMENTAL RESULTS AND DISCUSSION}
\subsection{The hybridized cavity-magnon mode in the linear range}

  The experimental setup is sketched in Fig.~\ref{fig:setup}(a). A polished YIG sphere with $1$~mm diameter is placed at the magnetic field node of a Fabry-Perot-like cavity, which is an assembled apparatus with a circular waveguide connecting to coaxial-rectangular adapters~\cite{Hyde2018Direct,Yao2015Theory}, such that the excited magnon and cavity photon are dissipatively coupled. The cavity is pumped by a microwave generator, and the transmission is detected by a signal analyzer. The embedded YIG sphere is placed in a static magnetic field $H$ produced by tunable electromagnets at room temperature, which is not depicted in Fig.~\ref{fig:setup}(a).
  
  We first conduct the experiment under linear conditions, using a vector network analyzer to measure the transmission of the cavity with input power below the threshold to create the nonlinear effect, so that the Kerr nonlinear effect is negligible. Our cavity resonance is at $\omega _c/2\pi=12.8888$~GHz. The resonance frequency of Kittel-mode in our experiments follows the dispersion $\omega _m =\gamma _0(H_r + H_A)$, where $\gamma _0/2\pi = 29.86 \mu _0$~GHz/T is the gyromagnetic ratio, $\mu _0H_A=-6.1$~mT is the anisotropy field, and $H_r$ is the biased static magnetic field at resonance. When the frequency of the Kittel-mode is tuned in resonance with the cavity microwave photons, the standard level attraction of the hybridized modes, which is the signature of dissipative coupling, was measured and is shown in Fig.~\ref{fig:setup}(b). On the left side of this level attraction, an additional mode split caused by the high order spin wave is not of immediate interest for the discussion of dissipatively coupled nonlinear bistable effect. The dispersion of the hybridized cavity mode and magnon mode is shown in Fig.~\ref{fig:setup}(c), where points A and E correspond to far off-resonance condition with $|\omega-\omega_c|\gg2\Gamma$, and point C indicates on-resonance condition with $|\omega-\omega_c|=0$, and points B and D indicate an intermediate frequency condition with $|\omega-\omega_c|\sim 2\Gamma$.  The dissipative coupling strength can be determined by the separated gap at $\omega _m=\omega _c$ in Fig.~\ref{fig:setup}(d), i.e., $\Gamma/2\pi=21$~MHz. The intrinsic and extrinsic linewidth of cavity-mode $\kappa _i/2\pi=2.58$~MHz, $\kappa _e/2\pi=126$~MHz are obtained by fitting the transmission coefficient spectra when $|\omega _c-\omega _m|\gg 2\Gamma$ where the coupling effect is negligible. As seen from Fig.~\ref{fig:setup}(d), the fitting agrees well with the experimental results. The intrinsic and extrinsic dampings of the magnon are $1.6$ MHz and $3.5$ MHz respectively. Hence, the total damping of magnon is $\gamma/2\pi=5.1$~MHz.
\subsection{Field foldover hysteresis loop}

\begin{figure}[t]
\includegraphics[width=0.5\textwidth]{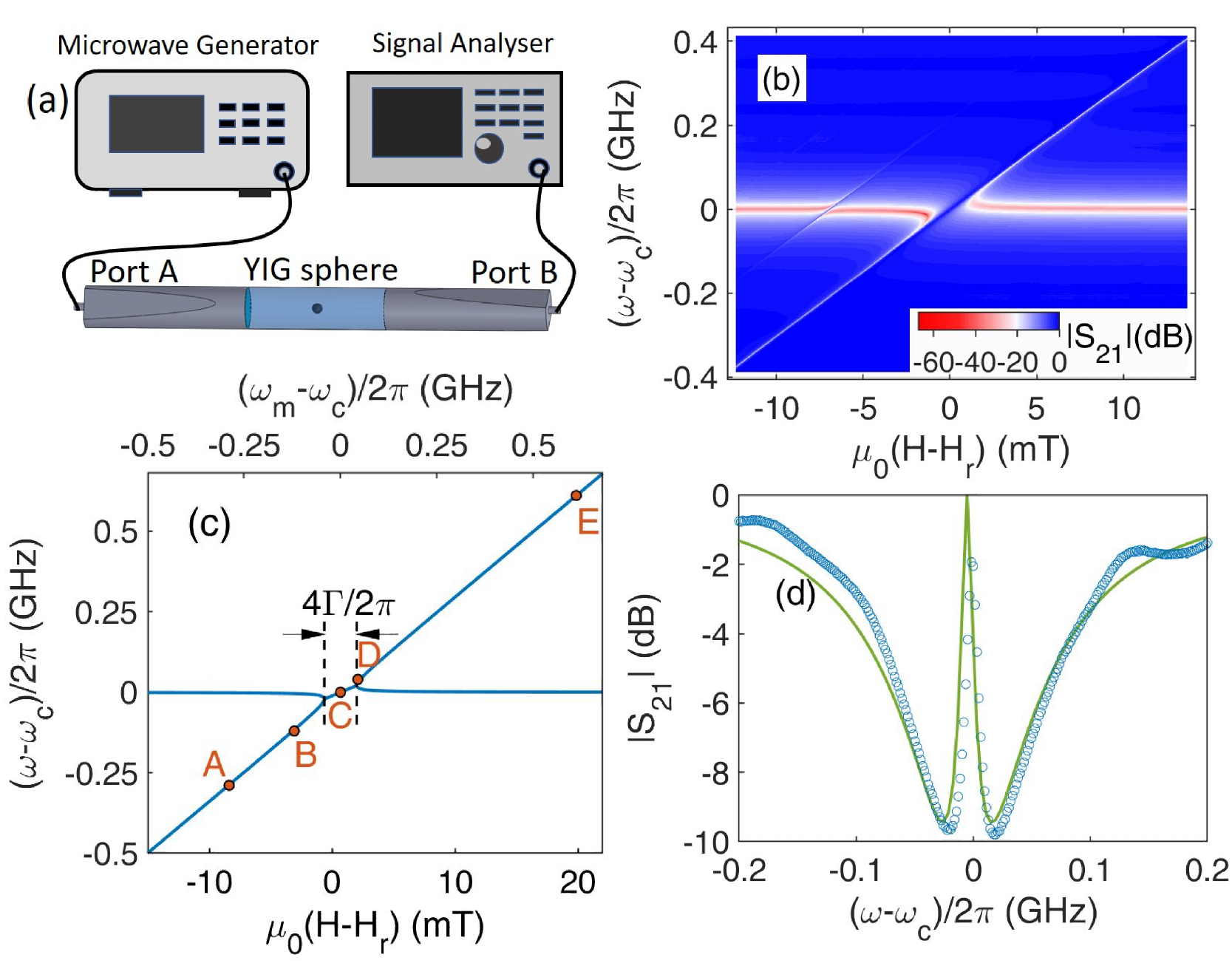} 
  \caption{\label{fig:setup}(a) Illustration of the experimental setup, where a YIG sphere is imbedded in an assembly cavity. Here, the YIG is at the node of microwave magnetic field. The cavity is pumped by the microwave generator and the transmission is measured by signal analyzer. (b) Transmission coefficient mapping of the hybridized cavity-magnon system, with level attraction implying dissipative coupling. White dashed line indicates the cut at the coupling point with $\omega _m=\omega _c$. (c) Dispersion relation of hybridized cavity and magnon mode, where points A and E indicate off-resonance with $|\omega-\omega_c|\gg2\Gamma$, and point C indicates on-resonance with $|\omega-\omega_c|=0$, and points B and D indicate intermediate frequencies with $|\omega-\omega_c|\sim 2\Gamma$. (d)~The fixed field cut of transmission coefficient mapping at the coupling condition $\omega _m=\omega _c$ indicated by white dashed line in (b), with symbols for the experimental data and solid line for calculation based on Eq.~(\ref{tansmission equation}).} 
\end{figure}

In this section, nonlinear effects in our coupled cavity-magnon system for on- and far off-resonance frequencies are measured by sweeping the magnetic field. Here high microwave powers provided by a microwave generator are used to drive the large angle precession of the magnon, while the transmission signal is measured by a signal analyzer, as depicted in Fig.~\ref{fig:setup}(a). 

\begin{figure}[b]
  \includegraphics[width=0.5\textwidth,trim=1 1 5 1,clip]{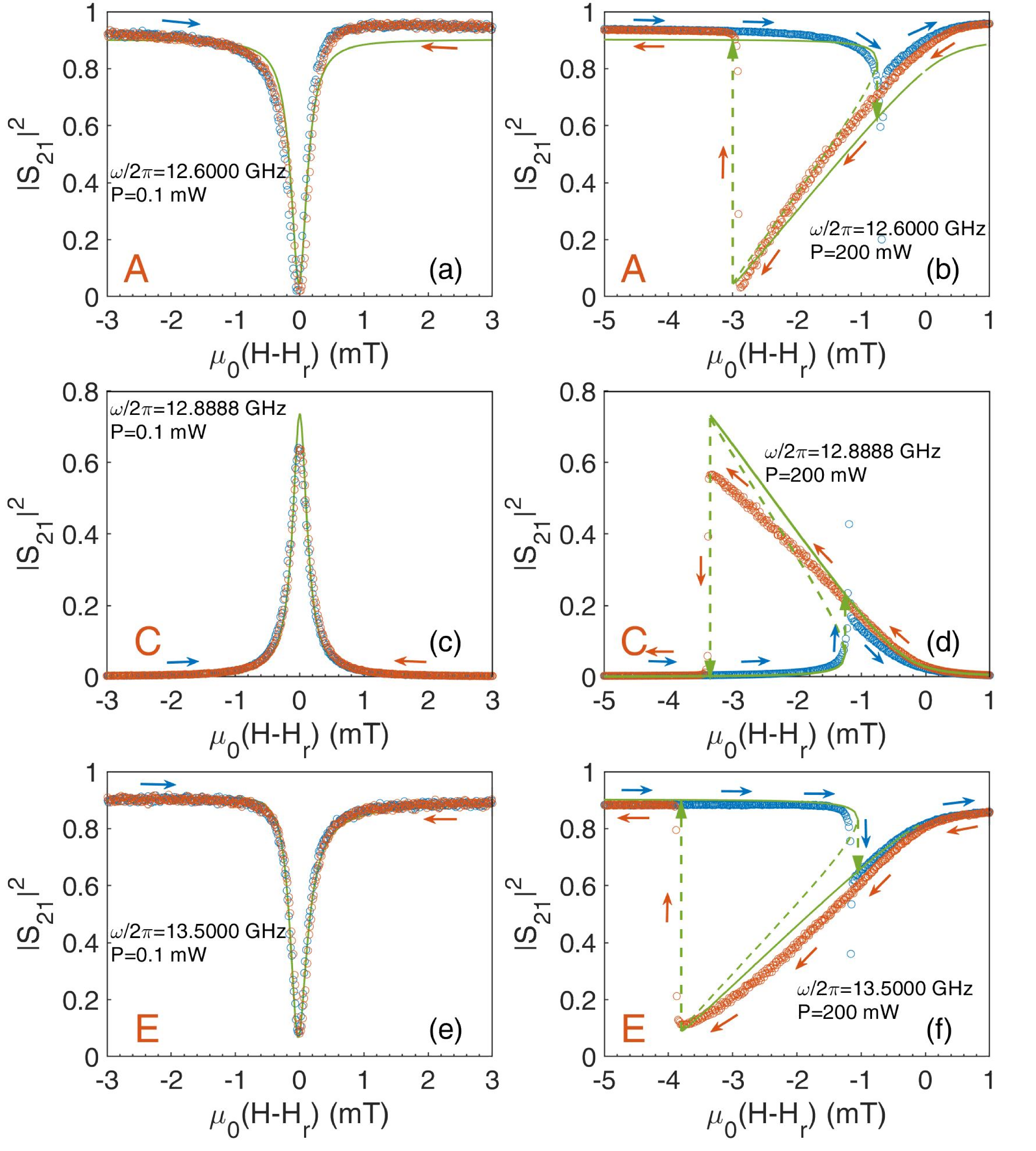}
  \caption{\label{fig:field foldover 1}$|S_{21}|^{2}$ versus H with (a)~$P=0.1$~mW and (b)~$P=200$~mW at off-resonant frequency $\omega/2\pi =12.6000$~GHz. $|S_{21}|^{2}$ versus $H$ with (c)~$P=0.1$ mW and (d)~$P=200$~mW at on-resonant frequency $\omega/2\pi =12.8888$~GHz. $|S_{21}|^{2}$ versus $H$ with (e) $P=0.1$~mW and (f) $P=200$~mW at off-resonant frequency $\omega/2\pi =13.5000$~GHz. Blue~(orange) circle symbols are experimental data by up-sweeping~(down-sweeping) static magnetic field $H$. The solid curves are calculated. Dashed green lines with arrows indicate the transition process of bistability and dashed green lines without arrows indicate the unstable state.} 
\end{figure}

We start our measurements in the linear range by setting the output power of the microwave generator to be $0.1$~mW. The transmission was measured at an on-resonance frequency $\omega /2\pi=12.8888$~GHz indicated by point C in Fig.~\ref{fig:field foldover 1}(c), and off-resonance frequencies $\omega /2\pi = 12.6000$~GHz, $13.5000$~GHz indicated by points A and E in Fig.~\ref{fig:field foldover 1}(c), when we perform up-sweeping and down-sweeping magnetic field, as shown in Figs.~\ref{fig:field foldover 1}(a), \ref{fig:field foldover 1}(c) and \ref{fig:field foldover 1}(e), respectively. At conditions $\omega /2\pi = 12.6000$~GHz and $\omega /2\pi =13.5000$~GHz, where the magnon mode is dominant, the spectra show a minimum transmission at the resonance condition~$H = H_r$ because of strong absorption due to the magnon excitation as shown in Figs.~\ref{fig:field foldover 1}(a) and \ref{fig:field foldover 1}(e). In contrast, the spectrum shows a maximum transmission at condition~$\omega=\omega _c$ as shown in Fig.~\ref{fig:field foldover 1}(c). This peak originates from two factors:~(1)~the cavity strongly absorbs microwaves at resonance near $\omega= \omega _c$ with the small transmission producing the flat background; (2)~when the cavity mode dissipatively couples to a YIG magnon mode at $\omega_m= \omega _c$, the hybrid system will thus result in a maximum transmission signal at $H = H_r$. The peak of transmission spectra indicates half-photon and half-magnon mode.

\begin{figure}[tb]
  \includegraphics[width=0.5\textwidth,trim=8 0 0 0,clip]{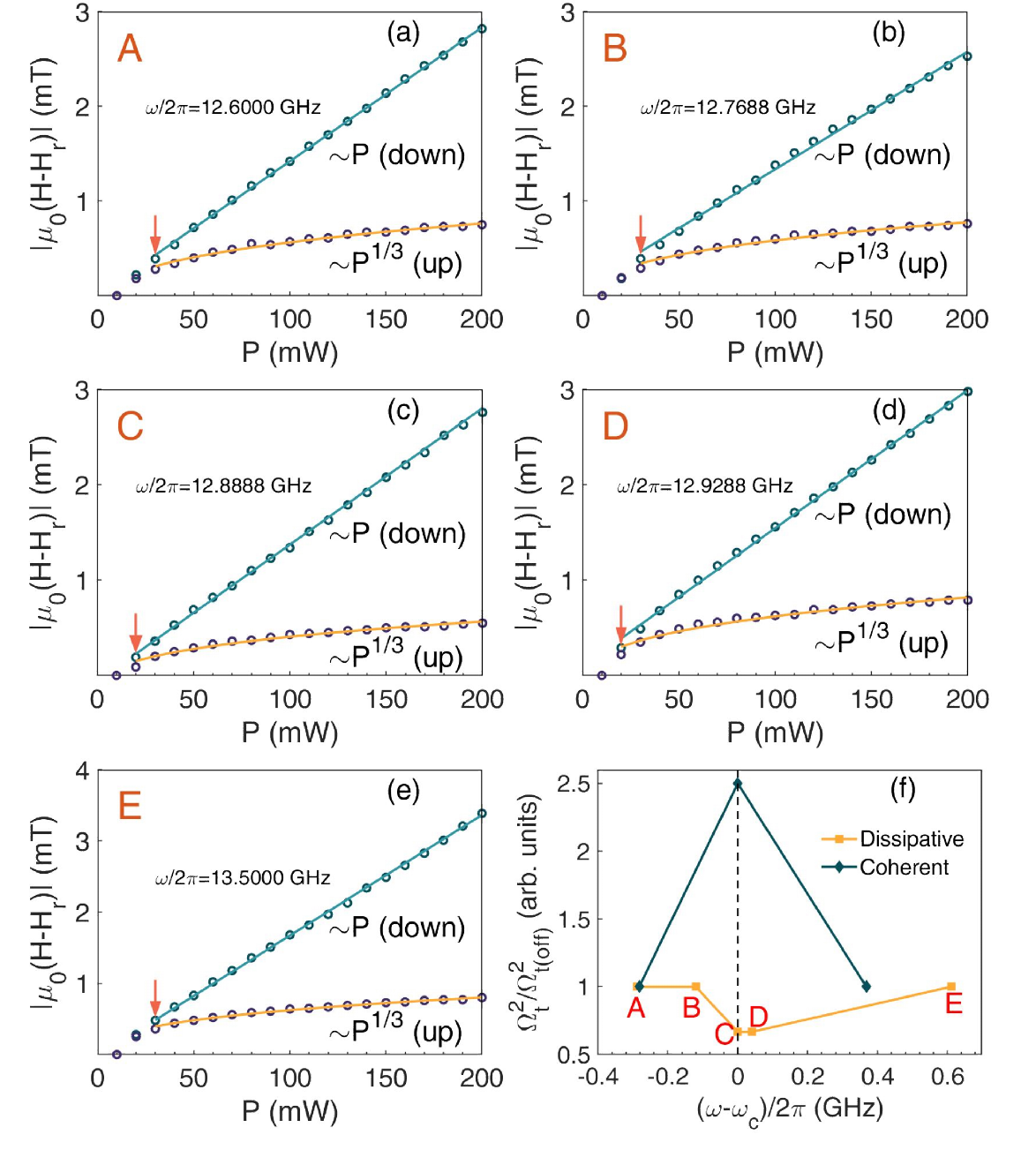}
  \caption{\label{fig:field transition of bistability}The jump position of field foldover hysteresis versus P at cavity frequencies (a) $\omega/2\pi=12.6000$~GHz,~(b)~$\omega/2\pi=12.7688$~GHz, (c) $\omega/2\pi=12.8888$~GHz,~(d) $\omega/2\pi=12.9288$~GHz, (e) $\omega/2\pi=13.5$~GHz. Purple (blue) circle symbols are experimental results of forward~(backward) H field sweeping. The yellow and cyan solid curves are fitted using Eqs.~(\ref{upper border}) and (\ref{down boarder}) for forward and backward sweeping, respectively. (f) The threshold for coherent and dissipative coupling system when off- and on-resonance. The threshold for coherent coupling is from Ref.~\cite{Hyde2018Direct}. For simplicity, we consider each threshold relative to that of off-resonance in each system. Here we have utilized the relation $\Omega_t^2/\Omega_{t(off)}^2=P_t/P_{t(off)}$, where $\Omega_{t(off)}$ and $P_{t(off)}$ are the critical driving field and power off-resonance, respectively.} 
\end{figure}

In order to study the nonlinear effect, we increase power above the threshold. By up- and down-sweeping the magnetic field, we observe that two abrupt jumps of transmission spectra occur at different static magnetic field biases $H$, corresponding to abrupt transitions between these two stable states. In the range of the transition, a hysteresis loop is clearly seen in the up- and down-sweeping traces of transmission spectra shown in Figs.~\ref{fig:field foldover 1}(b), \ref{fig:field foldover 1}(d) and \ref{fig:field foldover 1}(f). This behavior can be explained by Eq.~(\ref{nonlinear_b_power}), which predicts the behavior of bistability and transitions between the two stable states when the power is above the threshold. The hysteresis loop becomes more evident with the increasing power, because the transition points will depart from each other as microwave power increases, as shown in Fig.~\ref{fig:field transition of bistability}. 

The bistability of off- and on-resonance have distinctly different behaviors in our magnon-cavity system. When the system is off-resonance, i.e.,~$|\omega-\omega _c|\gg 2\Gamma$, the field hysteresis loops (Figs.~\ref{fig:field foldover 1}(b) and \ref{fig:field foldover 1}(f)) are clockwise when considering the up- and down-sweeping direction of the static magnetic field. In contrast, when the system is on-resonance, i.e., at $\omega = \omega _c$,  the field hysteresis loop in Fig.~\ref{fig:field foldover 1}(d) is counterclockwise. This behavior is quite different from the bistability of coherently coupled magnon-photon system as measured in Ref.~\cite{Hyde2018Direct}. The work in Ref.~\cite{Hyde2018Direct} demonstrated when $K<0$ there is clockwise hysteresis for on-resonance (with peak background) and anti-clockwise hysteresis for off-resonance~(with dip background).

The direction of hysteresis depends on the sign of $K$ and the background shape of the resonance. For any nonlinear mode with a Lorentzian dip or peak resonance that is excited with high power, the trace of the transmission spectrum will jump at the last transition point along the sweeping direction because of the hysteresis phenomena. For instance, suppose $K>0$, the trace of the transmission spectrum with peak background will jump at the right transition point from a low to high amplitude by up sweeping the magnetic field, while it will jump at the left transition point from a high to low amplitude by down sweeping the magnetic field. Thus, the hysteresis for $K>0$ and peak background lineshape is counterclockwise. By the same approach, the direction of hysteresis can be summarized in Table \ref{direction of hysteresis}, through which the difference in the direction of hysteresis among our measurement and the work of \cite{Wang2018Bistability,Hyde2018Direct} is well explained. 
 
\begin{table}[!htbp]
	\centering
	\caption{Direction of hysteresis.}
  \begin{ruledtabular}
	\begin{tabular}{lp{3cm}p{3cm}p{3cm}}
		~  ~~~~~~~~~~~~~~               &\textbf{peak}   &\textbf{dip}    \\
		\hline  
		$K>0$    & counterclockwise   & clockwise                    \\
		$K<0$     &clockwise       & counterclockwise \\
	\end{tabular}
\end{ruledtabular}
  \label{direction of hysteresis}
\end{table}

\begin{figure}[t]
  \includegraphics[width=0.5\textwidth,trim=0 0 5 5,clip]{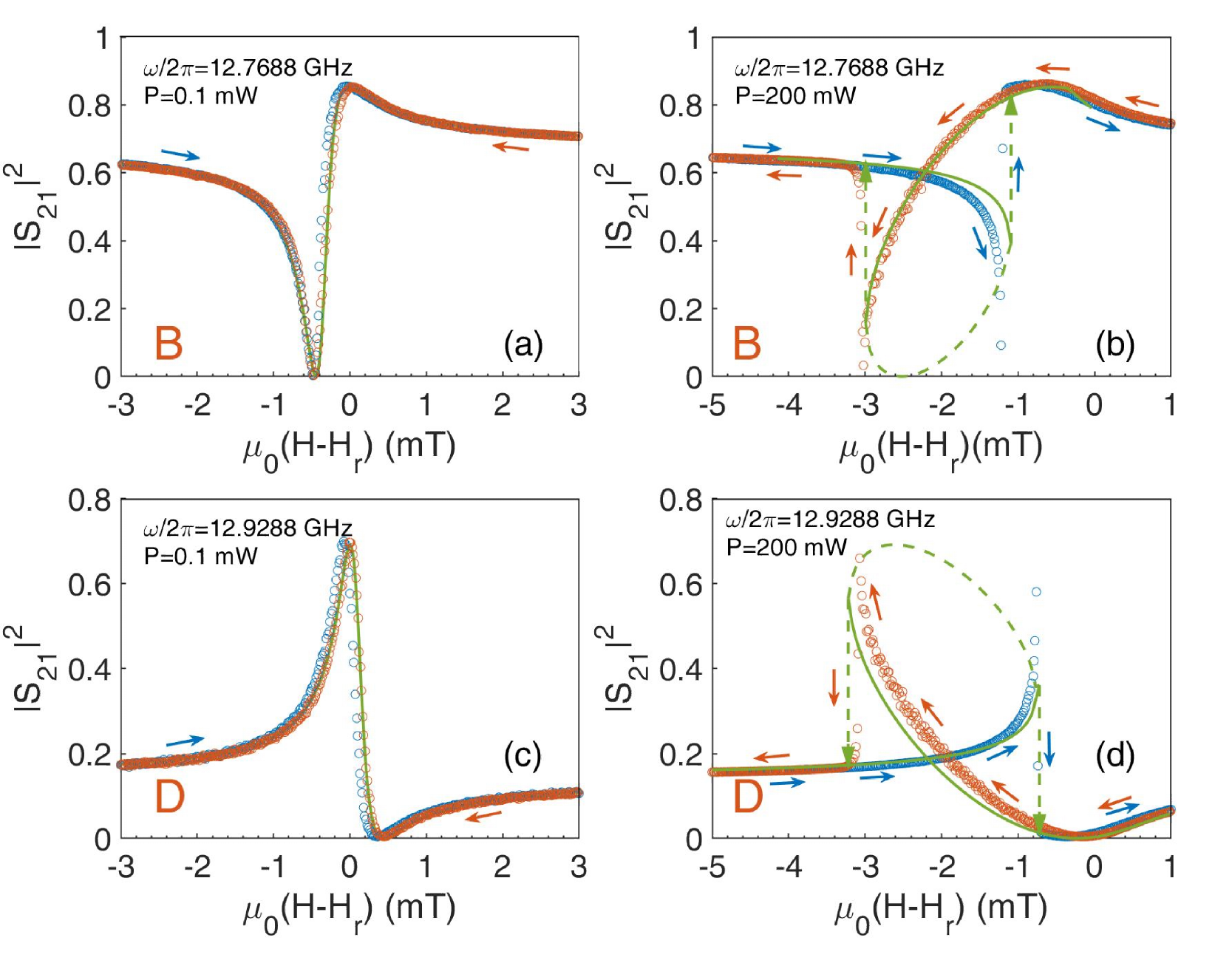}
  \caption{\label{fig:field foldover 2}$|S_{21}|^{2}$ versus H with (a)~$P=0.1$~mW and (b)~$P=200$~mW at intermediate frequency $\omega/2\pi =12.7688$ GHz. $|S_{21}|^{2}$ versus $H$ with (c)~$P=0.1 $~mW and (d)~$P=200 $~mW at intermediate frequency $\omega/2\pi =12.9288 $~GHz. Blue~(orange) circle symbols are experimental data by up-sweeping~(down-sweeping) the static magnetic field $H$. The solid curves are calculated. Dashed green lines with arrow indicate the transition process of bistability and dashed green lines without arrow indicate the unstable state.} 
\end{figure}

When $|\omega-\omega _c|\sim 2\Gamma$ indicated by points B and D shown in Fig.~\ref{fig:field foldover 1}(c), a different foldover hysteresis loop is seen at intermediate frequencies above and below $\omega _c$. Generally the lineshape of transmission spectrum is symmetric when the power is below the nonlinear threshold, for instance, a typical Lorentzian peak characteristic at $\omega = \omega _c$ and a Lorentzian dip $|\omega -\omega _c| >> 2\Gamma$. However, the lineshape of transmission spectrum is asymmetric when the frequency is tuned to the region among $|\omega-\omega _c|\sim 2\Gamma$, shown in Figs.~\ref{fig:field foldover 2}(a) and \ref{fig:field foldover 2}(c) with input power $0.1$~mW. This asymmetric lineshape, similar to that in the coherent scenario~\cite{Hyde2018Direct}, is due to Fano-like resonance~\cite{Fano1961Effects}. However, their polarities are opposite because of different coupling mechanism.
As microwave power is increased to $200$~mW, in contrast to the general hysteresis loop on- and far off-resonance, a butterfly-like hysteresis loop appears and the polarities of the butterfly-like hysteresis loop are opposite when the microwave frequency is set at intermediate frequencies above and below $\omega _c$ as shown in Figs.~\ref{fig:field foldover 2}(b) and \ref{fig:field foldover 2}(d). This difference of shape results from the transition direction of bistability. For on- and off-resonance frequency, the direction of two transitions are opposite, where one is from the low to high transmission, and another is from the high to low transmission. While, for the intermediate frequencies, the direction of two transitions are the identical, i.e., both from high to low transmission or from low to high transmission. 

 The effective dampings of the magnon are $\gamma_m'/2\pi=4.50,~0.20,~0.10,~0.15,~5.10$~MHz by fitting the transmission spectra when $P=0.1$~mW for different frequencies A-E, respectively. The fitted effective damping of the magnon reveals that the damping of the magnon is suppressed on-resonance due to the dissipative interaction between the cavity and the magnon, which is consistent with the result of Ref.~\cite{yao2019microscopic}. Then, with the damping parameters of the cavity and magnon extracted from the linear process and the solution of Eq.~(\ref{nonlinear_b_power}), we obtain the fitted parameter $KS^2=4.5\times 10^{-9},~6.2\times 10^{-8},~8.7\times 10^{-8},~9.9\times 10^{-8},~7.0\times 10^{-9}$~$\rm GHz^{3}/mW$ at cavity frequency $\omega/2\pi=12.6000,~12.7688,~12.8888,~12.9688,~13.5000$~GHz, respectively. (Here, $K$ and $S$ can not be determined individually.) In addition, the fitted transmission spectrum as a function of magnetic field can reproduce the field-sweeping bistability as shown with green line in Figs.~\ref{fig:field foldover 1} and \ref{fig:field foldover 2}. This agreement verifies the validity of our generalized model which describes dissipatively coupled Duffing oscillator and linear oscillator in this quasi-one-dimensional cavity~\cite{Yao2015Theory}.

\begin{figure*}[htbp]
  \includegraphics[width=1\textwidth,trim=0 39 16 40,clip]{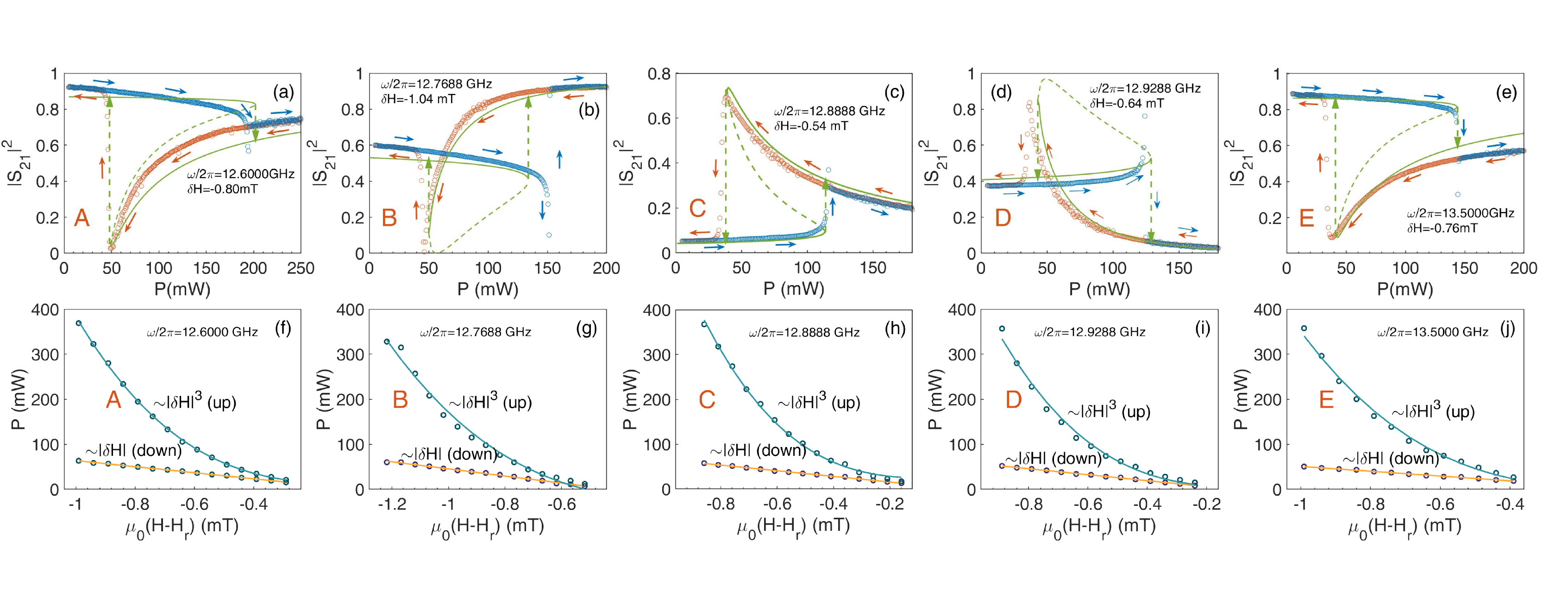}
  \caption{\label{fig:power bistability and critical field} The power foldover hysteresis with fixed static magnetic field at cavity frequencies (a)~$\omega/2\pi=12.6000$~GHz,~(b)~$\omega/2\pi=12.7688$~GHz,~(c)~$\omega/2\pi=12.8888$~GHz,~(d)~$\omega/2\pi=12.9288$~GHz,~(e)~$\omega/2\pi=13.5000$~GHz. The recorded power foldover hysteresis transition points versus magnetic field detuning $H-H_r$ at cavity frequencies (f)~$\omega/2\pi=12.6000$~GHz,~(g)~$\omega/2\pi=12.7688$~GHz,~(h)~$\omega/2\pi=12.8888$~GHz,~(i)~$\omega/2\pi=12.9288$~GHz,~(j)~$\omega/2\pi=13.5$~GHz.} 
\end{figure*}

\subsection{Power dependence of transition and threshold for field foldover hysteresis loop}
We have observed that the resonance gradually shifts toward lower $H$ when increasing the microwave power because the Kerr coefficient $K$ is positive~(see Appendix \ref{Hamiltonian of the coupled hybrid system}). While this is different from the negative Kerr coefficient system where the resonance gradually shifts toward high $H$~\cite{Hyde2018Direct}. When the power is above the threshold, the bistability appears, and its two transitions will shift depending on the power. Figures.~\ref{fig:field transition of bistability}(a)-\ref{fig:field transition of bistability}(e) show the jump positions as a function of the microwave power at $\omega/2\pi = 12.6000$, $12.7688$, $12.8888$, $12.9288$, and $13.5000$~GHz, respectively. As predicted by Eqs.~(\ref{upper border})  and (\ref{down boarder}), the up-sweeping transition (purple symbols) follows a $P^{1/3}$ dependence~(solid line) and the down-sweeping transition~(blue symbols) follows a linear $P$ dependence~(solid line) with fitted parameters $KS^2=4.3\times 10^{-9}$,~$5.8\times 10^{-8}$,~$8.9\times 10^{-8}$,~$8.7\times 10^{-8}$,~$6.5\times 10^{-9}$~$\rm GHz^{3}/mW$, respectively. This $KS^2$ magnitudes determined by fitting the transition points with the power dependence in Eq.~(\ref{down boarder}) is comparable to that fitted by transmission spectrum with Eq.~(\ref{tansmission equation}). The agreement of $KS^2$ magnitudes fitted by two different methods is gratifying in view of the error.

The power dependencies are different from those of a coherently coupled nonlinear system, where the down-sweeping jump follows a $P^{1/3}$ dependence and the up-sweeping jump follows a linear $P$ dependence~\cite{Hyde2018Direct}. Because the positive Kerr coeffcient corresponds with $\delta \omega_m<0$ and negative Kerr coeffcient corresponds with $\delta \omega_m>0$, these opposing frequency shifts will lead to the reversed power dependence of two transition points.

 Figures~\ref{fig:field transition of bistability}(a)-\ref{fig:field transition of bistability}(e) record the up-sweeping and down-sweeping transition point when increasing power for on- and off-resonance frequencies A-E. We can extract the threshold for each frequency and plot them in Fig.~\ref{fig:field transition of bistability}(f) for the dissipatively~[marked by orange arrows in Figs.~\ref{fig:field transition of bistability}(a)-\ref{fig:field transition of bistability}(e)] and coherently~(data from Ref.~\cite{Hyde2018Direct}) coupled systems with coupling strength $\Gamma/2\pi=21$~MHz, $g/2\pi=18$~MHz, respectively. The results in Fig.~\ref{fig:field transition of bistability}(f) reveal that the threshold of bistability on-resonance is about 2/3 of those off-resonance in a dissipatively coupled system. This bears stark disparities with a coherently coupled system, where the threshold of the cavity on-resonance is 2.5-fold of that of off-resonance in Ref.~\cite{Hyde2018Direct}. The results imply that the dissipative coupling indeed reduces the threshold despite the fact that our coupled system deviates from anti-PT conditions. In fact, the observed improvement of lower threshold originates from the threshold's cubic dependence on effective magnon damping which remains valid in the dissipative coupling case (see Table \ref{Threshold}) and the suppressed magnon damping in our dissipative coupling condition [(see Eq.~(\ref{damping of magnon})]. 

\subsection{Power foldover hysteresis loop and critical magnetic field for power bistability}
The hysteresis loops can also be observed by up-sweeping and down-sweeping power at fixed $\omega$ and $H$ as shown in the upper panels (a)-(e) of Fig.~\ref{fig:power bistability and critical field}, where we set the biasing magnetic field  $\delta H$ to be $-0.80$,~$-1.04$,~$-0.54$,~$-0.64$,~$-0.76$~mT at cavity frequency $\omega/2\pi=12.6000$, $2.7688$, $12.8888$, $12.9288$, $13.5000$~GHz, respectively. In contrast to the case of $K<0$ with opposite direction for power and field hysteresis~\cite{Hyde2018Direct}, here for $K>0$, they have identical direction at each frequency A-E as shown in Figs.~\ref{fig:field foldover 1}(b),~\ref{fig:field foldover 1}(d),~\ref{fig:field foldover 1}(f), \ref{fig:field foldover 2}(b),~\ref{fig:field foldover 2}(d), and \ref{fig:power bistability and critical field}(a)-\ref{fig:power bistability and critical field}(e).

The bottom panels (f)-(j) of Fig.~\ref{fig:power bistability and critical field} show that the transition points of power bistability have $|{\delta H}| ^{3}$ and $|\delta H|$ dependence for up-sweeping and down-sweeping power respectively, which agree with the theoretical prediction of Eqs.~(\ref{up power stability}) and (\ref{down}). As seen in Figs.~\ref{fig:power bistability and critical field}(f)-\ref{fig:power bistability and critical field}(j), the two transition points depart from each other, and the area of power hysteresis loops becomes larger when the bias magnetic field is tuned to be away from $H_r$, and vice versa. At a critical magnetic field $H_c$, the power hysteresis loops disappear. This phenomenon can be explained by Eq.~(\ref{critical condition for field}), which implies the requirement of producing power hysteresis loops is that the biasing magnetic field should be below the critical magnetic field at each frequency A-E.

\section{conclusions}
To summarize, we have observed both the field and power bistability of the dissipatively coupled cavity-magnon system. 
A theoretical model is studied in which a Duffing and linear oscillator are dissipatively coupled to explain the bistable behaviors. 
Such a dissipatively coupled hybridized system results in distinctly different bistable behaviors, like butterfly-like, clockwise, and counterclockwise hysteresis which are visualized through transmission spectra. 
 For the field bistability, the transition points show $P ^{1/3}$ and $P$  respective dependence when up-sweeping and down-sweeping the magnetic field. Correspondingly, for the power bistability, the transition points show $|{\delta H}| ^{3}$ and $|\delta H|$ dependence when up-sweeping and down-sweeping power, respectively. Meanwhile, the critical condition required for observing field and power bistability is obtained.
With the suppressed magnon damping and therefore lowered threshold for bistability, our system may lay the foundation for wide applications of very low power
nonlinearity devices. Besides, benefiting from flexible tunability with, e.g., the magnon frequency, the interaction strength between cavity and magnon, the drive power, the bistability of the cavity magnonics system may be potentially applied in emergent applications like memories and switches.

\begin{acknowledgments}
  This work has been funded by NSERC Discovery Grants and NSERC Discovery Accelerator Supplements~(C.-M. H.). Z.H. A. acknowledges the financial support from the National Natural Science Foundation of China under Grant Nos. 12027805/11991060, and the Shanghai Science and Technology Committee under Grant Nos. 20JC1414700, 20DZ1100604.~H. Pan was supported in part by the China Scholarship Council (CSC). The authors thank Garrett Kozyniak and Bentley Turner for discussions and suggestions.
\end{acknowledgments}
\appendix

\section{Hamiltonian of the coupled hybrid system}
\label{Hamiltonian of the coupled hybrid system}
The cavity-magnon hybrid system shown in Fig.~\ref{fig:setup}(a) includes a small YIG sphere with Kerr nonlinearity which is dissipatively coupled to a Fabry-Perot-like cavity, and the cavity is driven by a microwave field. The Hamiltonian of such a system consists of four parts (setting $\hbar\equiv 1$):
\begin{equation}
  H_{tot}=H_c+H_m+H_I+H_d,
\end{equation}
here $H_c =\omega_c \alpha^+\alpha$ corresponds to the bare Hamiltonian of the cavity mode, with  the creation (annihilation) operator $\alpha^+$~$(\alpha)$ at frequency $\omega_c$.

In our experiment, we apply a uniform static magnetic field $H = H\mathbf{e_z}$ orientating along the z-axis, and the YIG sphere has volume $V_m$. The static magnetic field is used to align the magnetization and tune the frequency of the magnon mode. When Zeeman energy and magnetocrystalline anisotropy energy are included, the Hamiltonian of the YIG sphere reads: 
\begin{equation}
  H_m=-\mu_0 \int_{V_m} \mathbf{M}\cdot \mathbf{H} d\tau-\frac{\mu_0}{2}\int_{V_m} \textbf{M}\cdot \mathbf{H_{an}}d\tau, 
  \label{magnon Hamiltonian}
\end{equation}
where $\mu_0$ is the vacuum magnetic permeability, $\mathbf{M} = (M_x,M_y,M_z)$ is the macrospin magnetization of the YIG sphere, and $\mathbf{H_{an}}$ is the magnetocrystalline anisotropy field in the YIG crystal. We have neglected the contribution of demagnetization energy  of the YIG sphere in Eq.~(\ref{magnon Hamiltonian}) as it is a constant term~\cite{blundell2001magnetism,Wang2016Magnon}.

For a uniformly magnetized YIG sphere, which is magnetized along $z$-axis with its anisotropy field along $z$-axis in our experiment, the anisotropy field can be written as $\mathbf{H_{an}} = mM_z\mathbf{e_z}$, where $m$ is dependent upon the dominant first-order anisotropy constant and the saturation magnetization~\cite{prabhakar2009spin}. Since $H_A<0$ in our experiment, we can obtain that $m<0$. Thus, the Hamiltonian of Eq.~(\ref{magnon Hamiltonian}) turns out to be
\begin{equation}
  H_m=-\mu_0H M_z V_m+\frac{1}{2}\mu_0 m M_z^2V_m.
  \label{magnon Hamiltonian M_z}
\end{equation}
Since the relation between macrospin magnetization $\mathbf{M}$ and macrospin operator $\mathbf{S}$~\cite{Wang2016Magnon,Wang2018Bistability,soykal2010strong} is 
\begin{equation}
  \mathbf{M}=-\frac{\gamma_0 \mathbf{S}}{V_m}=-\frac{\gamma_0}{V_m}(S_x,S_y,S_z),
  \label{relation of M and S}
\end{equation}
where $\gamma_0$ is the gyromagnetic ratio. By inserting such relation indicated by Eq.~(\ref{relation of M and S}) into Eq.~(\ref{magnon Hamiltonian M_z}), we obtain
\begin{equation}
  H_m=-\mu_0\gamma_0 H S_z-\frac{\mu_0m\gamma_0^2S_z^2}{2V_m}.
  \label{magnon Hamiltonian S_z}
\end{equation}
The first term of the above equation corresponds to Zeeman energy. The macrospin operators and the magnon operators are related via the Holstein-Primakoff transformation~\cite{holstein1940field}
\begin{eqnarray}
S^+&=&(\sqrt{(2S-b^+b)})b, \label{S+}   \\
S^-&=&b^+(\sqrt{(2S-b^+b)}),\label{S-}   \\
S_z&=&S-b^+b, \label{S,b}
\end{eqnarray}
where $S$ is the total spin number of the YIG sphere, $b^+$~$(b)$ the creation~(annihilation) operator of the magnon at frequency $\omega_m$, and $S^{\pm}\equiv S_x \pm iS_y$ are the raising and lowering operators of the macrospin. Through inserting Eq.~(\ref{S,b}) into Eq.~(\ref{magnon Hamiltonian S_z}), the Hamiltonian $H_m$ can be written as 
\begin{equation}
  H_m=\omega_m b^+b+Kb^+bb^+b,
\end{equation}
here $\omega_m=\mu_0\gamma_0 H+\mu_0\gamma_0^2mS/{V_m}$ denotes the frequency of the magnon mode and $K=-\mu_0\gamma_0^2m/{(2V_m)}$ the Kerr coefficient. Since $m<0$ for our experiment, the Kerr coefficient K is positive. Notably, the Kerr effect of magnons term $Kb^+bb^+b$ arises from magnetocrystalline anisotropy.
The  Hamiltonian representing the interaction between the magnon and the cavity mode is
\begin{equation}
  H_I=i\Gamma(b^++b)(a^++a),
\end{equation}
where $\Gamma$ denotes the dissipative coupling strength between the magnon and the cavity mode. With the rotating-wave approximation, we can neglect the fast oscillating terms~\cite{walls2008quantum}, and the cavity-magnon interaction Hamiltonian can be reduced as
\begin{equation}
  H_I=i\Gamma(b^+a+ba^+).
\end{equation}
 The interaction between the cavity photon and the drive field can be expressed as~\cite{agarwal2010electromagnetically}
\begin{equation}
  H_d=\Omega _d(a^+e^{-i\omega t}+ae^{i\omega t}),
\end{equation}
where $\Omega_d$ is the amplitude of the driving field.
Finally, we have the total Hamiltonian of the nonlinear cavity magnonics system where the cavity and magnon are dissipatively coupled, and the cavity is directly pumped 
\begin{eqnarray}
  &&\hspace{-1.5em}H_{tot} = {\omega _c}{a^ + }a + {\omega _m}{b^ + }b + K{b^ + }b{b^ + }b + i\Gamma \left( {{a^ + }b + a{b^ + }} \right)  \nonumber\\
  &&\hspace{-1.5em}~~~~~~~~~~~~~~~~~~~~+{\Omega _d}\left( {{a^ + }{e^{ - i\omega t}} + a{e^{i\omega t}}} \right).
\end{eqnarray}
\nocite{*}

\bibliography{draft_appendix}
\end{document}